# WST, the Wide-field Spectroscopic Telescope: Mechanical Design and FE Analyses for the High Resolution Spectrograph

Simone D'Auria[a], Andrea Tozzi[a], Anna Brucalassi[a], Matteo Munari[b], Ciro Del Vecchio[a], Maria Sofia Randich[a], Roland Bacon[c], David Lee[d],

[a]*INAF - Osservatorio Astrofisico di Arcetri, Largo Enrico Fermi 5, 50125 Firenze, Italy*

[b]*INAF - Osservatorio Astronomico di Catania, Via S. Sofia 78, 95123 Catania, Italy*

[c]*Centre de Recherche Astrophysique de Lyon, France*

[d]*STFC – UK Astronomy Technology Centre*

## ABSTRACT

The Wide-field Spectroscopic Telescope (WST) is a planned 12-meter-class dedicated spectroscopic facility designed to address key scientific challenges through large spectroscopic surveys. This paper presents the current status of Work Package 4.5, which focuses on the High-Resolution Multi-Object Spectrograph (MOS-HR) module for WST. The MOS-HR instrument is expected to provide a resolving power of R = 40,000 with a multiplexing capability of about 2,000 targets. The mechanical design activities carried out for the development of the HR spectrograph and for the definition of its opto-mechanical architecture are described. To account for both the scientific requirements of the spectrograph and the manufacturability constraints associated with such a complex instrument, the mechanical layout has been organized into four larger modules, each containing two sub-modules. Guided by feasibility considerations, such as mechanical performance, available volume, and fabrication and assembly aspects, each sub-module adopts a vertical optical bench configuration with optical elements mounted on both sides. Starting from the baseline optical design, the mechanical configuration has been developed to achieve the required alignment accuracy, structural stability, and environmental robustness. The workflow includes the translation of the optical prescription into a complete mechanical model, the definition of the main mounting and alignment interfaces, and preliminary static, modal, and seismic analyses to evaluate performance under operational and survival loads. As an outcome, the proposed design provides architecture that enables preliminary estimates of mass, volume, cost, and mechanical performance in terms of deformation, stress, and modal behavior of the modules.

## 1. INTRODUCTION

The Wide-field Spectroscopic Telescope (WST) [1] [2] is a next-generation 12-meter-class facility designed to address key scientific challenges through large spectroscopic surveys. The telescope is conceived to provide simultaneous Multi-Object Spectroscopy (MOS) [3] and Integral Field Spectroscopy (IFS) [4] capabilities over a wide field of view, operating in both Low- and High-Resolution modes [5] [6]. Within the WST instrumentation suite, the High-Resolution Multi-Object Spectrograph (MOS-HR) is intended to deliver a resolving power of $R = 40{,}000$ while simultaneously observing approximately 2,000 targets. Such demanding scientific and operational requirements impose significant constraints on the mechanical architecture of the instrument in terms of structural stability, alignment accuracy, manufacturability, and integration complexity. This work presents the actual status of the preliminary mechanical design activities carried out for the MOS-HR spectrograph within Work Package 4.5 of the WST project. Starting from the baseline optical prescription, the opto-mechanical architecture of the spectrograph has been developed with the goal of defining a feasible and scalable mechanical configuration compatible with the instrument requirements and available integration constraints. To improve manufacturability, assembly, and handling [7], the spectrograph has been organized into four main modules, each composed of two sub-modules adopting a vertical optical bench configuration with optical components mounted on both sides. This paper describes the development workflow adopted for the definition of the mechanical layout, including the translation of the optical design into a complete three-dimensional mechanical model, the identification of the main mounting and alignment interfaces, and the preliminary assessment of the structural behavior of the system through Finite Element Analyses (FEA). Static, modal, and seismic analyses have been performed to evaluate the response of the proposed configuration under operational and survival load conditions, providing initial estimates of stiffness, deformation, stress distribution, and modal characteristics. The results obtained represent a first step toward the consolidation of the MOS-HR opto-mechanical architecture and the refinement of subsequent detailed design phases.

## 2. Mechanical Design and Opto-Mechanical Architecture

### 2.1 Optical Design Overview

The baseline optical design of the MOS-HR spectrograph is based on a reflective collimator architecture composed of a class 1-meter spherical mirror coupled with an aspherical corrector plate. To reduce the overall footprint of the instrument, the optical path is folded through a flat mirror tilted at 45 degrees. To allow the beam to emerge from the pseudo-slit to reach the collimator mirror, the folding mirror includes a central aperture accommodating the slit assembly.

The pseudo-slit geometry is approximately cylindrical, with a length of about 266 mm and a thickness between 2 and 5 mm. Fibers are arranged linearly along the slit, with a preliminary pitch distance of 100 μm and a core diameter of 73 μm. The adopted configuration foresees the use of 7 sliced fibers per target, with an additional empty pitch introduced to reduce spectral crosstalk.

After collimation, the beam is spectrally separated through dichroic elements and dispersed by transmissive gratings operating at approximately 45 degrees incidence angle, allowing the overall system volume to remain compact.

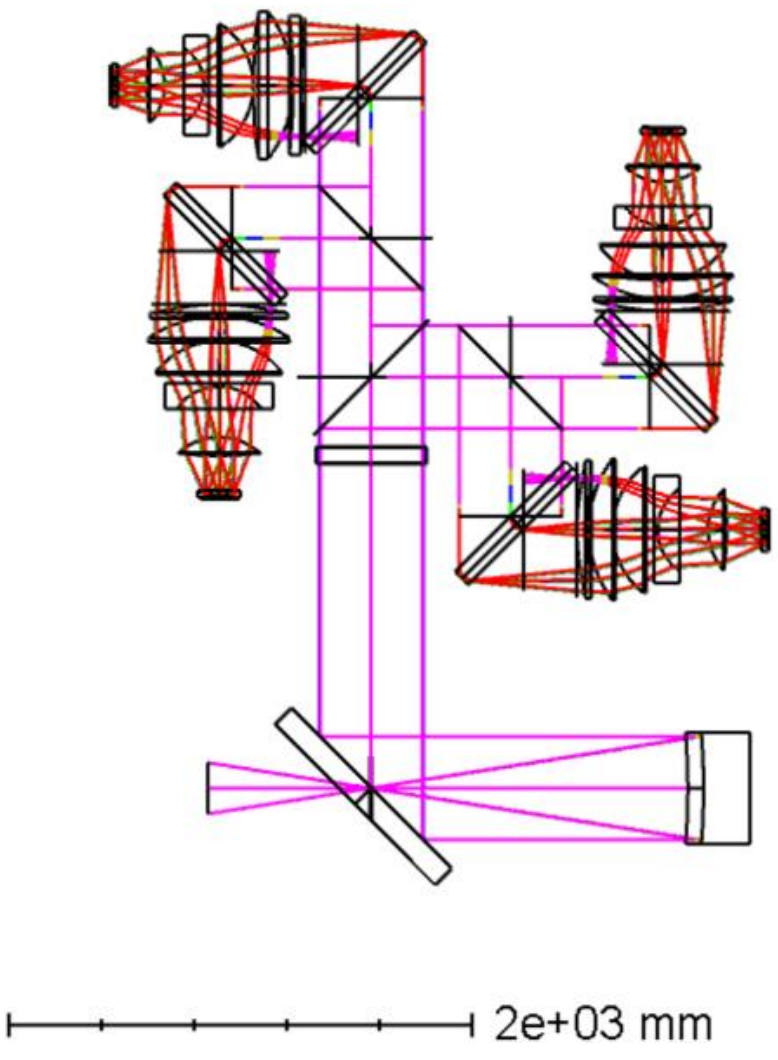

*Figure 1 - Optical Design of the MOS-HR*

Based on the current slit configuration and fiber packing assumptions, each spectrograph module can accommodate approximately 330 simultaneous objects. Multiple replicated modules are therefore required to achieve the overall multiplexing capability targeted by the MOS-HR instrument. Preliminary estimates indicate that the combined optical and mechanical footprint of a single module is approximately 18 m², making compactness, modularity, and manufacturability key aspects of the mechanical design process.

### 2.2 Mechanical Design Drivers

The optical configuration described above introduces several constraints that strongly influence the mechanical architecture of the spectrograph. The large dimensions of the optical components, the elongated pseudo-slit geometry, the compact folded optical path, and the requirement to replicate multiple identical modules demand a mechanical solution capable of combining structural stiffness, compactness, manufacturability, and ease of integration.

The design activity has therefore been focused on reducing the overall instrument footprint while maintaining adequate structural stability and limiting relative displacements between optical elements. At the same time, particular care has been devoted to the accessibility of alignment interfaces and to simplifying integration and maintenance procedures. Additional constraints arise from the need to ensure the manufacturability and transportability of the individual modules, as well as to accommodate detector assemblies and associated cryogenic interfaces within a highly compact layout.

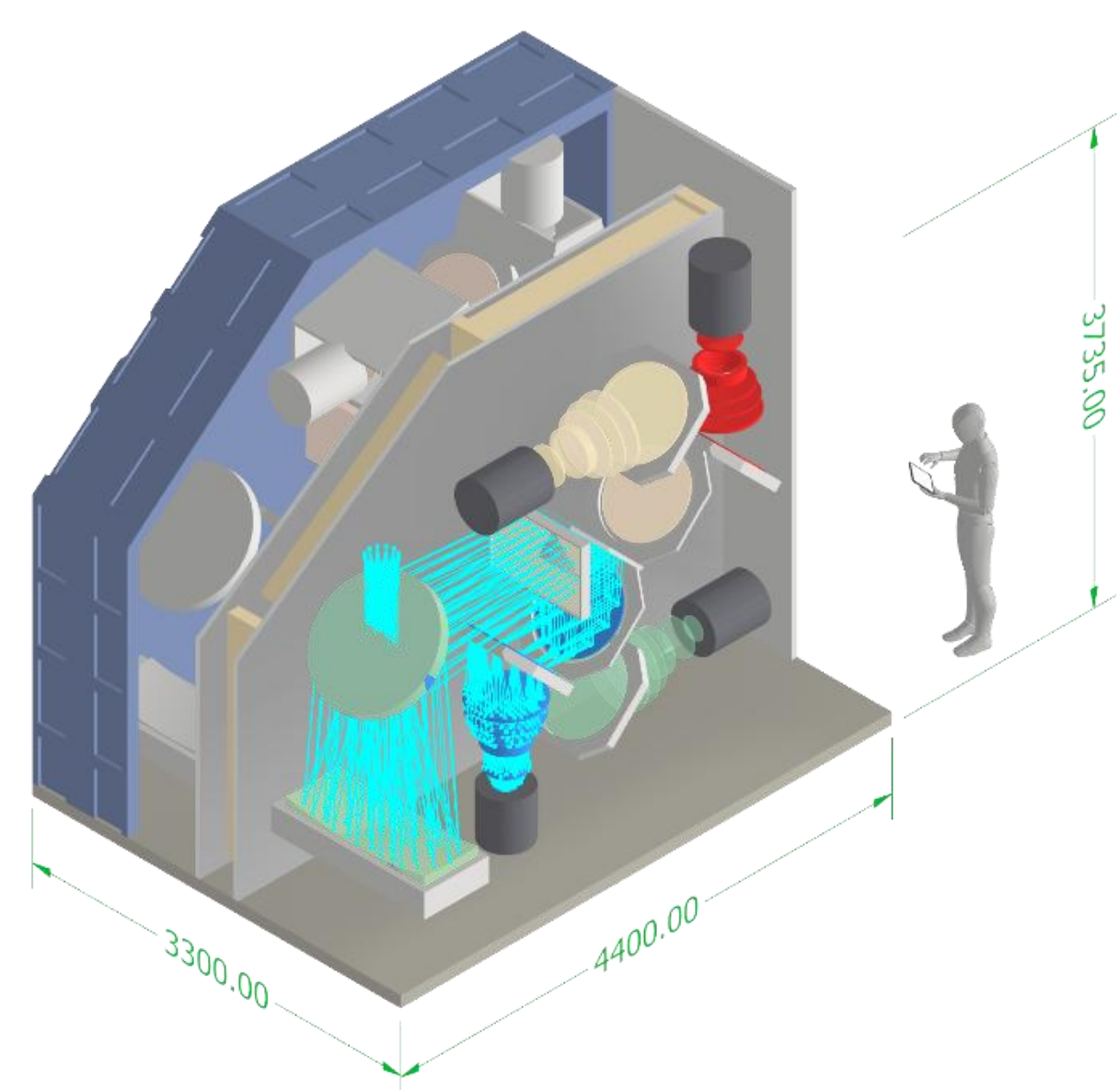

*Figure 2 - Preliminary design of the MOS-HR double module*

These considerations guided the development of a modular opto-mechanical architecture based on vertically oriented optical benches, with optical components mounted on both sides of the supporting structure. This configuration was selected as a compromise between structural performance, volume optimization, manufacturability, and assembly requirements.

# 3. OPTO-MECHANICAL ARCHITECTURE DEVELOPMENT

## 3.1 Modular Architecture Concept

The mechanical architecture of the MOS-HR spectrograph has been developed starting from the need to combine High Multiplexing capability with a compact and manufacturable instrument configuration. Since the required multiplexing can only be achieved through the replication of multiple spectrograph units, modularity rapidly emerged as one of the main design drivers during the preliminary development phase.

The proposed architecture foresees the subdivision of the instrument into multiple independent modules, each integrating two spectrographs within a common mechanical structure. This approach allows the reduction of the overall footprint on the Telescope rotating platform while maintaining a manageable envelope for manufacturing, transportation, and assembly. In addition, the modular configuration simplifies possible future maintenance and replacement operations by limiting the intervention to individual sub-systems [8].

The adopted philosophy follows an approach conceptually similar to that implemented in existing highly multiplexed spectroscopic instruments such as MOONS [9], where replication and compact packaging represent key enabling factors for achieving large multiplexing capabilities within constrained Telescope environments. In the case of MOS-HR, the compactness requirement becomes particularly relevant due to the large number of optical assemblies required by the instrument baseline design.

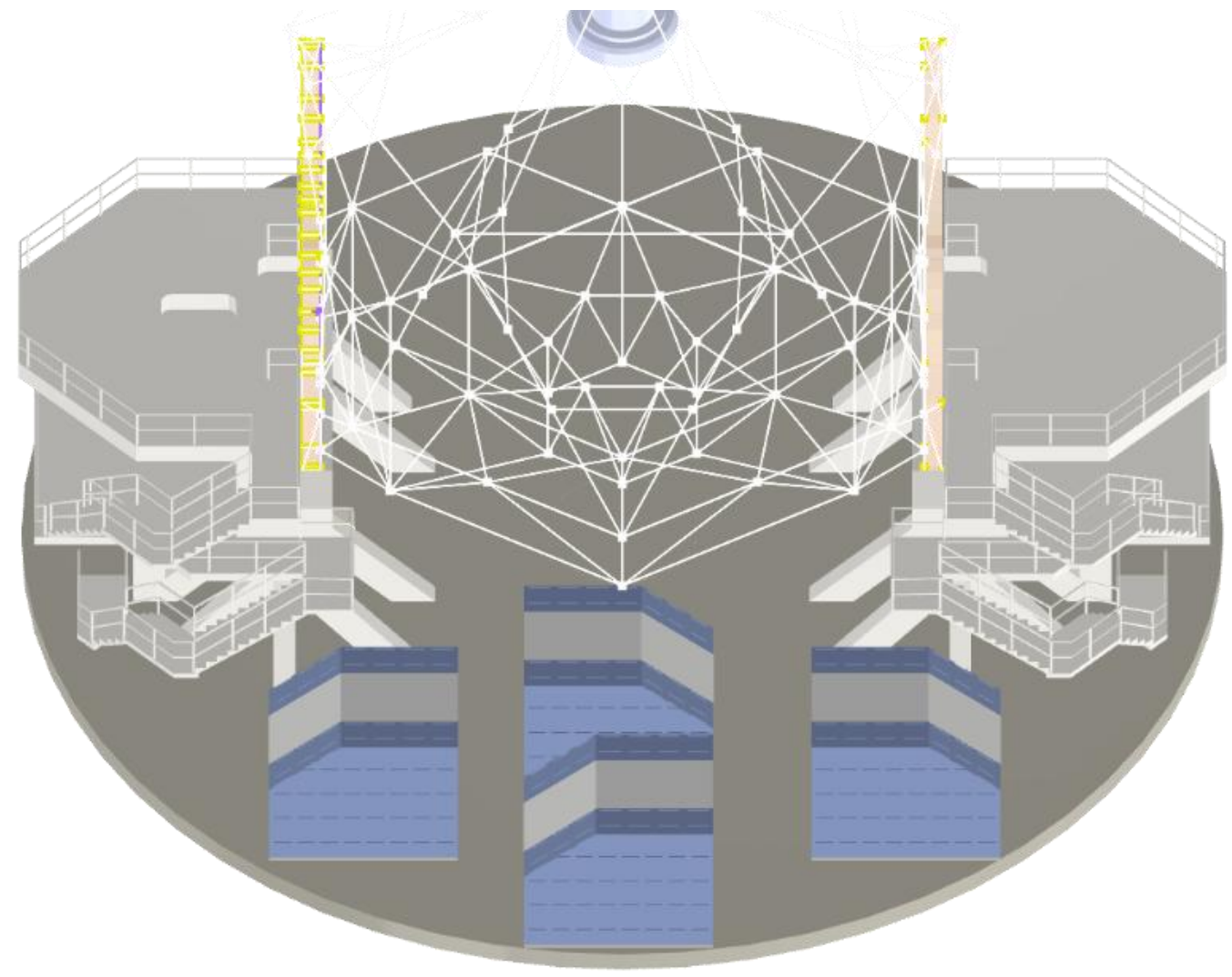

*Figure 3 - Possible MOS-HR layout of 4 double modules on the rotating platform of the WST*

## 3.2 Vertical Optical Bench Configuration and Structure Design

To optimize the available volume and reduce the footprint on the telescope rotating platform, the spectrograph modules have been developed around a vertically oriented optical bench configuration. Within each module, the optical components are distributed on both sides of the supporting structure, allowing a more compact layout compared to a conventional horizontal arrangement while also improving accessibility to interfaces, detectors, cabling, and alignment areas.

The optical bench has been preliminarily conceived as a lightweight stiffened structure composed of two external skins connected through an internal rib network. This configuration was selected to provide a good compromise between bending and torsional stiffness, structural mass, manufacturability, and integration accessibility. The distribution of optical assemblies on opposite sides of the bench also contributes to a more balanced mass allocation and to a more efficient use of the available internal volume.

Attention has been devoted to the central region of the spectrograph, where the folded optical path introduces one of the most critical opto-mechanical interfaces of the instrument. The folding mirror, positioned at approximately 45 degrees with respect to the incoming beam, includes a central aperture allowing the pseudo-slit assembly to pass through the mirror itself. This solution makes it possible to accommodate the elongated slit geometry while maintaining a relatively compact instrument envelope. A preliminary concept for the slit assembly has also been developed, including first implementations of the fiber slit, shutter mechanism, and fiber back-illumination system required for calibration and alignment procedures.

The preliminary sizing of the optical bench has been driven by estimates of the masses associated with the optical components. Starting from the nominal optical masses, additional contributions related to mounts, interfaces, support structures, and auxiliary systems have been introduced using conservative assumptions based on similar existing instruments and preliminary engineering considerations. These global mass estimates have then been used to define the initial dimensions and thicknesses of the structural elements composing the bench, with particular attention devoted to limiting the maximum deflection under operational loading conditions and maintaining compatibility with the expected alignment stability requirements of the spectrograph.

### 3.3 Preliminary Integration Considerations

From the early stages of the design process, the mechanical layout has been developed, taking into account assembly, integration, and accessibility requirements. The modular subdivision of the instrument allows individual spectrograph units to be assembled and verified independently before final integration on the telescope platform.

The vertical optical bench configuration also facilitates access to optical interfaces during alignment activities, particularly in the central region of the spectrograph, where the highest concentration of optical and mechanical interfaces is present. Detector systems, slit assemblies, and auxiliary sub-systems can be accessed from dedicated sides of the module, reducing interference between integration procedures and simplifying maintenance operations.

Given the limited volume available in the central region of the spectrograph, different configurations have been explored for the integration of the slit assembly and its auxiliary subsystems. In the current concept, the pseudo-slit is positioned within the central aperture of the folding mirror, minimizing the overall instrument envelope and keeping the optical path compact. A possible solution under investigation foresees the integration of a compact shutter mechanism directly within the slit assembly. Besides its primary function during calibration procedures, the shutter could also host a fiber back-illumination system, combining multiple functionalities within a single subsystem and reducing the complexity of the surrounding interfaces. Although still at a conceptual stage, this approach appears promising from both integration and compactness perspectives.

Although still preliminary, the current architecture provides a coherent framework for the future refinement of alignment strategies, handling procedures, transportation configurations, and interface standardization activities required for the subsequent development phases of the MOS-HR instrument.

# 4. PRELIMINARY STRUCTURAL ASSESSMENT

## 4.1 Finite Element Model and Assumptions

A simplified Finite Element (FE) model of the optical bench has been developed to obtain a first-order estimate of the structural response of the MOS-HR module under representative loading conditions. The model is based on the double-skin and internal rib architecture described in the previous section and includes the main load-bearing elements of the vertical optical bench.

For this preliminary assessment, the structure has been assumed to be entirely manufactured in Aluminum Alloy, consistent with common practices in large astronomical instrumentation where a compromise between stiffness, mass, and manufacturability is required. The optical bench is considered rigidly constrained at its base, representing the interface with the supporting structure of the telescope platform.

The applied loads are derived directly from the estimated masses of the optical components introduced in the design phase. These masses are applied as lumped or distributed contributions on both sides of the optical bench skins to reproduce the asymmetric and distributed nature of the optical payload. In addition to the nominal optical masses, the model includes a first-order estimate of the associated mechanical interfaces and support structures, obtained by applying conservative scaling factors based on existing instruments with comparable complexity and size.

This approach allows the definition of a global load case that, although simplified, is considered sufficiently representative for preliminary evaluation of the structural behavior of the system.

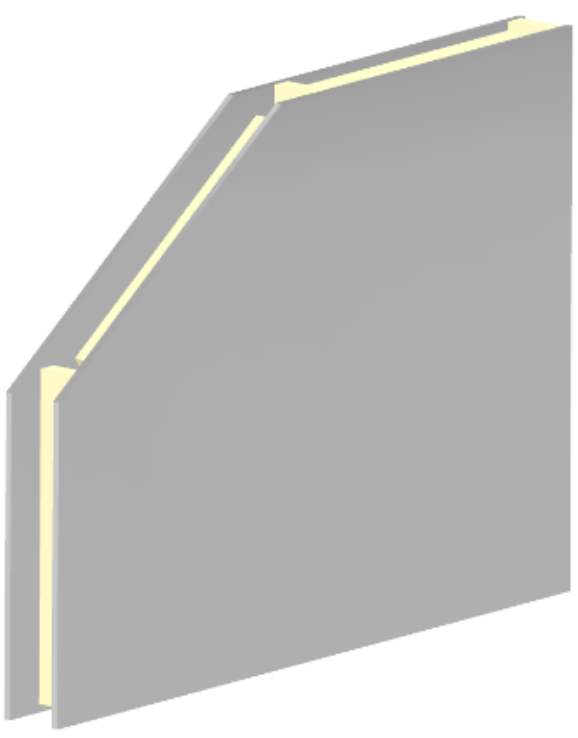

*Figure 4 - Vertical bench made of two skins with internal ribs*

## 4.2 Static Structural Response

The FE model has been developed through the COMSOL Multiphysics ® software [10]. The static analysis has been performed to evaluate the global deformation of the optical bench under gravity loading conditions. Attention has been devoted to the deflection of the two external skins and to the resulting relative displacements between opposite sides of the structure, as these directly affect the alignment stability of the optical system.

The results provide a first-order indication of the stiffness of the proposed double-skin configuration with internal ribs. The load distribution induced by optical components mounted on both sides of the bench generates a combined bending and torsional response of the structure, which is mitigated by the presence of the internal rib network that ensures the load transfer between the two skins.

The maximum deformation values obtained in this preliminary configuration are used as an initial benchmark to verify the compatibility of the current design with the expected alignment tolerances of the spectrograph. Although no detailed optical tolerance budget is applied at this stage, the results confirm that the proposed structural concept is consistent with the required order of magnitude of mechanical stability.

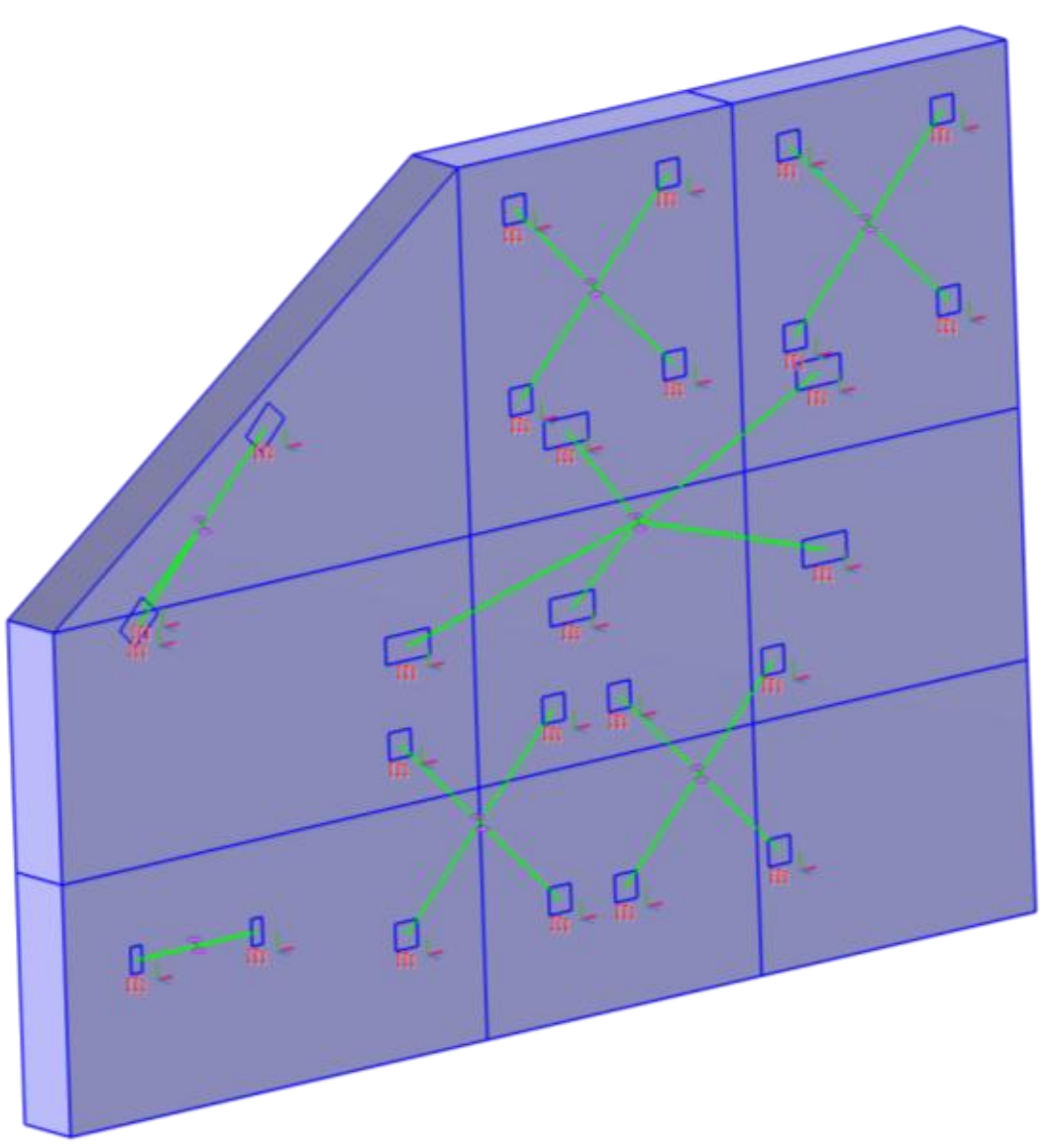

*Figure 5 - Vertical optical bench FEA geometry and setup: shell elements for both skin and internal ribs elements. Green: mass and moment of inertia of the optical element and mechanics estimated to be mounted on both skins*

### 4.3 Modal Behavior and Frequency Estimates

A preliminary modal analysis has been carried out to estimate the dynamic behavior of the optical bench and to obtain an order-of-magnitude evaluation of the fundamental resonance frequencies of the system. The same simplified FE model used for static analysis has been employed, maintaining identical boundary conditions and mass assumptions.

The first eigenmodes are dominated by global bending and torsional deformations of the vertical optical bench, reflecting the distributed nature of the optical payload on both sides of the structure. The resulting fundamental frequencies provide an initial indication of the overall dynamic stiffness of the system and are used as a reference for future design iterations and for comparison with potential environmental excitation sources associated with telescope operation.

Although still preliminary, the obtained frequency range is considered suitable for a first feasibility assessment of the structural concept. The results highlight the importance of the internal rib configuration and of the double-skin architecture in increasing the global stiffness of the module.

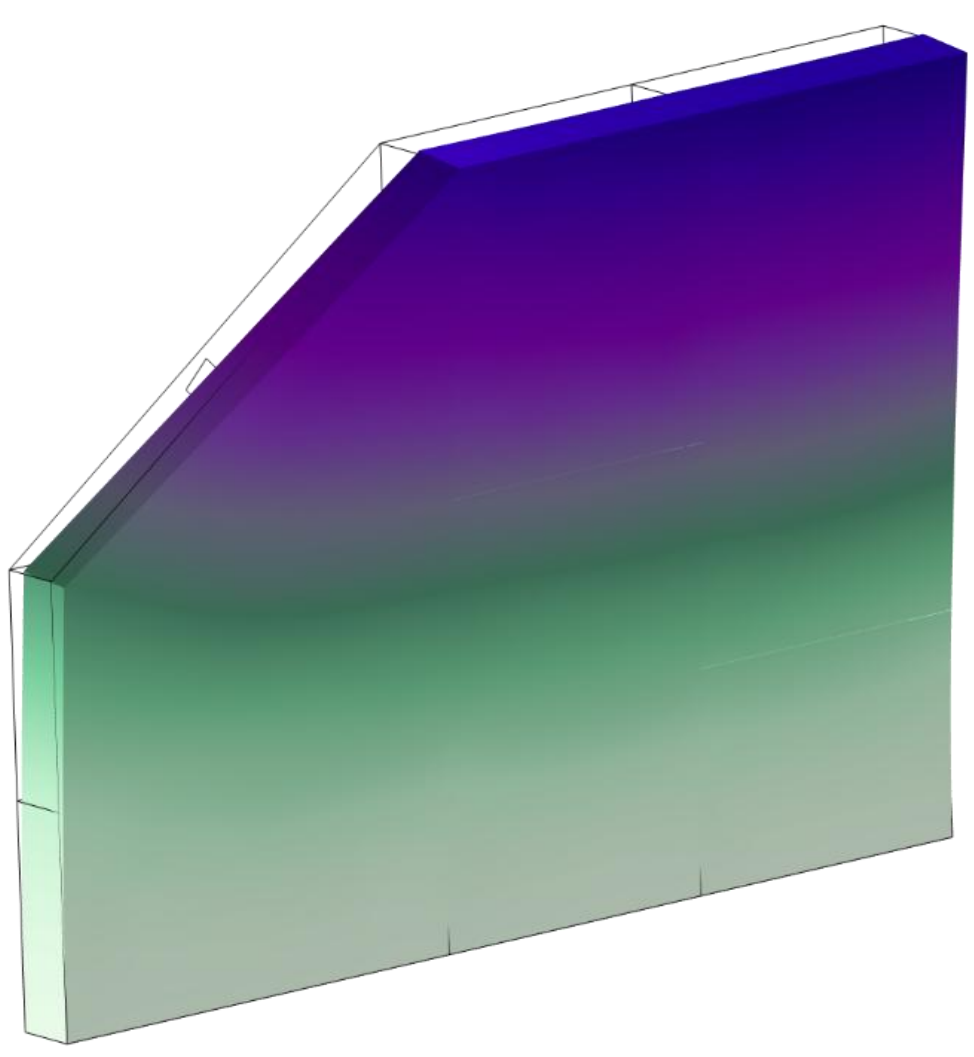

*Figure 6 - First eigenfrequency modal shape of the optical bench at around 10 Hz. BC: fixed at the bottom, masses and moments of inertia estimated for each optical and mechanical element*

### 4.4 Discussion and Design Implications

The combined static and modal analyses provide a first quantitative validation of the proposed opto-mechanical architecture of the MOS-HR spectrograph. Despite the simplifying assumptions adopted in the model, the results are considered sufficiently representative to support the current design choices and to guide subsequent optimization activities.

In particular, the analysis confirms the relevance of accurately modelling the contribution of optical masses and associated mechanical interfaces, which represent the dominant loading condition for the optical bench. The adopted approach, based on nominal optical masses complemented by conservative estimates of mechanical contributions derived from comparable instruments, provides a reasonable and cautious basis for preliminary structural evaluation.

These results constitute a starting point for more detailed analyses in future design phases, including refined interface modelling, thermo-mechanical effects, and higher-fidelity dynamic simulations aimed at fully characterizing the performance of the MOS-HR opto-mechanical system.

## 5. CONCLUSION AND FUTURE WORK

This paper has presented the preliminary mechanical design of the High-Resolution Multi-Object Spectrograph (MOS-HR) for the Wide-field Spectroscopic Telescope (WST), with focus on the translation of the baseline optical design into a coherent opto-mechanical architecture and on a first, rather simplified, structural assessment.

Starting from the optical layout, which is characterized by a compact folded path, a quite long pseudo-slit, and a high multiplexing requirement, the main mechanical drivers have been identified. These include strong constraints on stiffness, footprint on the rotating platform, modularity, and manufacturability aspects. Based on these needs, a modular concept has been defined, with spectrograph units grouping two optical channels per module, in order to keep the overall system more compact and still manageable in terms of integration.

A key aspect of the design is the vertical optical bench configuration, where optical components are placed on both sides of a double-skin structure with internal ribs. This solution was primarily driven by the need to reduce the instrument's projected footprint while maintaining reasonable structural behavior and allowing access to optical interfaces. The design also includes initial concepts for critical areas, such as the folded optical path with the slit passing through the folding mirror, and early ideas for the fiber shutter and back-illumination used for calibration.

A simplified FE model of the optical bench has been developed in order to run preliminary static and modal checks. The structure has been loaded using mass estimates coming from the optical design, with additional conservative assumptions for mechanical parts, mostly based on similar instruments found in literature or experience. The results give a first idea of global deformations and natural frequencies, and overall, they look in line with what is expected for this kind of instrument, at least at this stage of the design.

Future work will mainly go in the direction of refining both the opto-mechanical layout and the structural model. More detailed interface definitions between optics and structure will be needed to better capture real mass and stiffness distribution. Thermal and thermo-mechanical effects should also be included in the next steps, together with more realistic boundary conditions closer to the telescope environment, in order to improve the dynamic response estimation.

From a design point of view, further optimization of the modular architecture and of the internal optical bench is still expected, especially regarding mass reduction, stiffness improvement, and simplification of assembly. A more complete tolerance and alignment strategy will also be important to consolidate the final mechanical design.

Overall, this work sets a first consistent baseline for the MOS-HR opto-mechanical design and provides a starting point for the next, more detailed, design and verification phases.

## ACKNOWLEDGMENTS

This project has received funding from the European Union’s Horizon Europe research and innovation programme under grant agreement No. 101183153.